# Spin-orbit-entangled state of $Ba_2CaOsO_6$ studied by O $K$-edge resonant inelastic X-ray scattering and Raman spectroscopy

J. Okamoto,[1] G. Shibata,[2] Yu. S. Ponosov,[3] H. Hayashi,[4,5,*] K. Yamaura,[4,5] H. Y. Huang,[1] A. Singh,[1] C. T. Chen,[1] A. Tanaka,[6] S. V. Streltsov,[3,7,†] D. J. Huang,[1,8,9,‡] and A. Fujimori[1,2,10,11,§]

[1]*National Synchrotron Radiation Research Center, Hsinchu 300092, Taiwan*
[2]*Materials Sciences Research Center, Japan Atomic Energy Agency, Sayo, Hyogo 679-5148, Japan*
[3]*Institute of Metal Physics, 620041 Ekaterinburg GSP-170, Russia*
[4]*Research Center for Materials Nanoarchitectonics (MANA), National Institute for Materials Science, Tsukuba, Ibaraki 305-0044,Japan*
[5]*Graduate School of Chemical Sciences and Engineering, Hokkaido University, Sapporo, Hokkaido 060-0810, Japan*
[6]*Department of Quantum Matter, Hiroshima University, Hiroshima 739-8530, Japan*
[7]*Department of Theoretical Physics and Applied Mathematics, Ural Federal University, 620002 Ekaterinburg, Russia*
[8]*Department of Physics, National Tsing Hua University, Hsinchu 300044, Taiwan*
[9]*Department of Electrophysics, National Yang Ming Chiao Tung University, Hsinchu 300093, Taiwan*
[10]*Center for Quantum Science and Technology and Department of Physics, National Tsing Hua University, Hsinchu 300044, Taiwan*
[11]*Department of Physics, The University of Tokyo, Bunkyo-Ku, Tokyo 113-0033, Japan*

Transition-metal ions with $5d^2$ electronic configuration in a cubic crystal field are prone to have a vanishing dipolar magnetic moment but finite higher-order multipolar moments, and they are expected to exhibit exotic physical properties. Through an investigation using resonant inelastic X-ray scattering (RIXS), Raman spectroscopy, and theoretical ligand-field (LF) multiplet and *ab initio* calculations, we fully characterized the local electronic structure of $Ba_2CaOsO_6$, particularly, the crystal-field symmetry of the $5d^2$ electrons in this anomalous material. The low-energy multiplet excitations from RIXS at the oxygen $K$ edge and Raman-active phonons both show no splitting. These findings are consistent with the ground state of Os ions dominated by magnetic octupoles. Obtained parameters pave the way for further realistic microscopic studies of this highly unusual class of materials, advancing our understanding of spin-orbit physics in systems with higher-order multipoles.

## I. INTRODUCTION

$5d$ transition-metal oxides exhibit various exotic physical properties arising from the strong spin-orbit coupling (SOC) that competes with Hund's coupling and Jahn-Teller effect and strongly influences the exchange interaction [1–3]. Particularly attractive have been the Mott-insulating $Ir^{4+}$ ($5d^5$) compounds with effective angular momentum $J_{\mathrm{eff}} = 1/2$ and the Kitaev quantum-spin-liquid candidate of $Ru^{3+}$ ($4d^5$) honeycomb lattices. Especially exotic are localized $5d^2$ electrons in a cubic crystal field. A $d^2$ ion coordinated by ligand atoms in the octahedral ($O_h$) environment is expected to be Jahn-Teller active, but many cubic crystals with $5d^2$ ions remain undistorted, probably due to the strong SOC of the $5d$ electrons [4]. Such a $5d^2$ ion in the $O_h$-symmetry crystal field has a non-Kramers doublet ground state that supports either an electric quadrupole or a magnetic octupole [5–9]. This is contrasted with $5d^1$ systems, where the $5d$ ion has an electric quadrupole and Jahn-Teller distortion is induced [1, 2, 4, 7, 10]. As a staggered octupolar order has been predicted theoretically for quarter-filled manganites [11], one may conceive disordered octupoles or a quantum octupole liquid under particular conditions in manganites or other materials.

The $B$-site-ordered double perovskite $Ba_2AOsO_6$, where $A$ is an alkali-earth metal, is one of such $5d^2$ ($Os^{6+}$) systems. The face-centered cubic (fcc) lattice formed by the Os ions may induce geometrical frustrations between the multipoles, and may lead to intriguing quantum magnetism predicted theoretically [5–9]. For example, $Ba_2CaOsO_6$ shows a cusp-like anomaly in the magnetic susceptibility at $T^* \sim 50$ K, signaling a magnetic transition, and muon-spin rotation ($\mu$-SR) has revealed a small magnetic moment of $\sim 0.2\ \mu_B$ [12] or $\sim 0.05\ \mu_B$ per Os ion [13] whereas neutron scattering has detected no magnetic Bragg peaks below $T^*$ [12]. (According to Ref.[13], the small magnetic moment may be induced by impurities and may not be intrinsic.) According to X-ray diffraction, the crystal remains cubic down to the lowest temperatures [14]. This precludes static electric quadrupolar order, which should distort the cubic lattice [10], but is consistent with magnetic octupolar order as the origin of the 'hidden order' in $Ba_2AOsO_6$. Theo-retically, exchange coupling between neighboring Os ions favors ferro-octupolar order in the fcc lattice [6].

* present address: *Institute for Solid State Physics, The University of Tokyo, 5-1-5 Kashiwanoha, Kashiwa, Chiba 277-8581, Japan*
† email: *streltsov@imp.uran.ru*
‡ email: *djhuang@nsrrc.org.tw*
§ email: *fujimori@phys.s.u-tokyo.ac.jp*

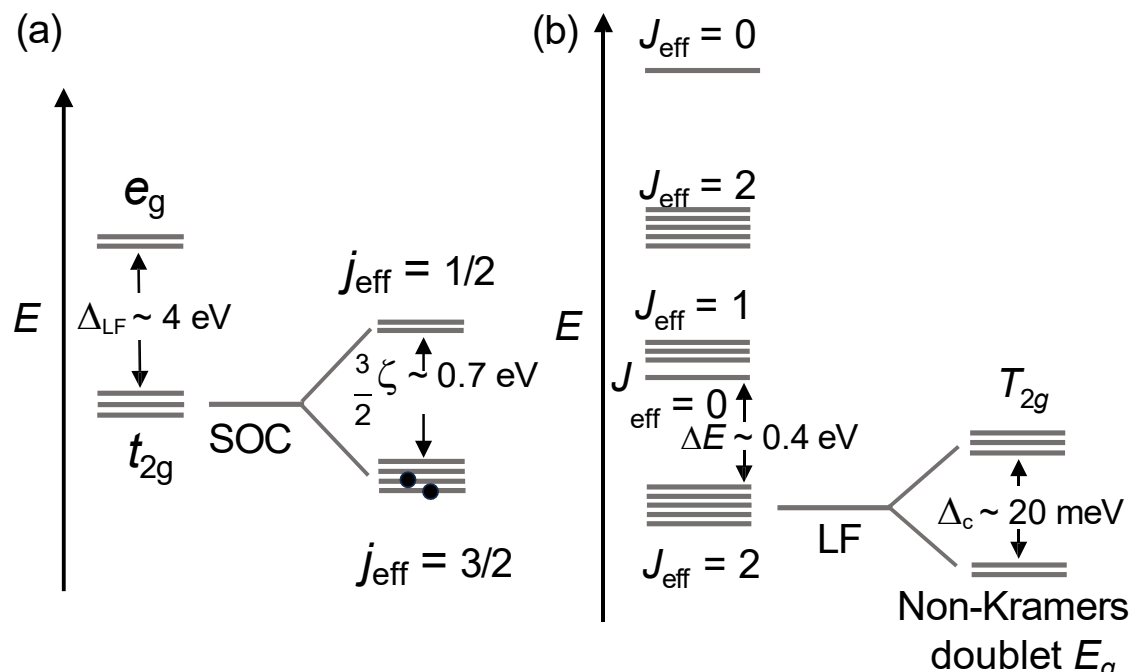

FIG. 1. Energy diagrams of the $Os^{6+}$ ($5d^2$) ion in the cubic ($O_h$) crystal field. (a) One-electron energy diagram. $\Delta_{LF}$ is the $t_{2g}$-$e_g$ splitting. Spin-orbit coupling (SOC) further splits the $t_{2g}$ level into $j_{eff} = \frac{1}{2}$ and $j_{eff} = \frac{3}{2}$ levels separated by $\frac{3}{2}\zeta$ and the $j_{eff} = \frac{3}{2}$ level is occupied by two electrons. (b) Two-electron energy diagram. The $t_{2g}^2$ part of the $d^2$ multiplet is shown whereas the $t_{2g}e_g$ part is located at higher energies separated by ~ $\Delta_{LF}$. Due to the ligand field (LF) of cubic symmetry, the $J_{eff} = 2$ ground state is split by a 'residual' cubic splitting $\Delta_c$ into the non-Kramers $E_g$ doublet and the $T_{2g}$ triplet. The first $J_{eff} = 0$ excited state appears $\Delta E$ ~ 0.4 eV above the ground state, and the first $J_{eff} = 1$ excited state ~ 0.065 eV above it.

In an $O_h$-symmetry crystal field, the one-electron Os $5d$ level is split into the $t_{2g}$ and $e_g$ levels separated by $\Delta_{LF}$ [Fig. 1(a)]. The strong SOC splits the $t_{2g}$ level into the $j_{eff} = \frac{1}{2}$ and $j_{eff} = \frac{3}{2}$ sublevels, and the latter is occupied by the two electrons of the $Os^{6+}$ ion. In the two-electron energy diagram [Fig. 1(b)], the ground state has the total effective angular momentum of $J_{eff} = 2$. Under the $O_h$ symmetry, the 'residual' cubic crystal-field splits the $J_{eff} = 2$ quintet into the ground-state $E_g$ doublet and the triply-degenerate $T_{2g}$ excited states with a separation $\Delta_c$ [14, 15]. The $E_g$ ground state is a non-Kramers doublet consisting of $|\psi_{g,\uparrow}\rangle \equiv |J^z_{eff} = 0\rangle$ and $|\psi_{g,\downarrow}\rangle \equiv \frac{1}{\sqrt{2}}(|J^z_{eff} = 2\rangle + |J^z_{eff} = -2\rangle)$, whereas the $T_{2g}$ excited states consist of $|\psi_{e,\pm}\rangle \equiv |J^z_{eff} = \pm 1\rangle$ and $|\psi_{e,0}\rangle \equiv \frac{1}{\sqrt{2}}(|J^z_{eff} = 2\rangle - |J^z_{eff} = -2\rangle)$. In the ferro-octupole-ordered state, all the Os ions are in one of the two eigenstates, $|\psi_{g,\pm}\rangle \equiv \frac{1}{\sqrt{2}}(|\psi_{g,\uparrow}\rangle \pm i|\psi_{g,\downarrow}\rangle)$, of the octupole operator $T_{xyz} \propto \overline{J^x J^y J^z}$, where the overline denotes symmetrization.

While the magnetic properties of $Ba_2AOsO_6$ double perovskites have been widely studied and microscopic models to explain observed anomalies have been proposed, their electronic structure remains unexplored experimentally. The present paper aims to fill this gap. In particular, we studied $Ba_2CaOsO_6$ by X-ray absorption spectroscopy (XAS) and resonant inelastic X-ray scattering (RIXS) at the O $K$ edge. RIXS studies of $5d$ transition-metal oxides have so far been performed mainly at the transition-metal $L_{2,3}$ edge since one can directly study the spin and orbital excitation of the $5d$ states [16–19]. However, owing to the strong SOC of the $5d$ electrons, RIXS at the O $K$ edge can also be used to study spin excitations [Fig. 1(b)][20–24]. O $K$-edge RIXS has the advantage of having higher energy resolution than transition-metal $L_{2,3}$-edge RIXS, allowing us to study low-energy electronic excitation and electron-phonon interaction. We have also utilized Raman scattering to detect possible local lattice distortion that induces low-symmetry crystal fields. None of the above measurements have indeed shown evidence for the lowering of the cubic symmetry, favoring the scenario that the Os ions have dominantly octupolar moments in $Ba_2CaOsO_6$.

## II. RESULTS AND ANALYSES

We performed O $K$-edge RIXS on high-quality polycrystalline samples with the energy resolution of ~30 meV using $\pi$ polarized X-rays and 90° scattering angle (see Methods). The O $K$-edge XAS spectrum is shown in Fig. 2(a). Figure 2(b) shows the RIXS intensity map in the $E_{in}$-$E_{loss}$ plane, where $E_{in}$ is the energy of incident X-ray and $E_{loss}$ is the energy loss of scattered X-rays. The same data are plotted in the $E_{in}$-$E_{em}$ plane in Fig. 2(c), where $E_{em}$ is the energy of emitted X-ray. RIXS spectra are plotted in Fig. 2(d). In the figure, above $E_{in}$ ~ 529 eV, some spectral features start to shift to higher $E_{loss}$ with increasing $E_{in}$, indicating a cross-over from Raman-like to fluorescence-like.

### Splitting due to ligand field and spin-orbit coupling

In O $K$-edge RIXS, the excitation of the O $1s$ core electron into empty states followed by the electron transition from the non-bonding O $2p$ band to the O $1s$ core level leaves a hole in the non-bonding O $2p$ band and an electron in the empty states. The resulting final state is equivalent to that of the O $2p$ → Os $5d$ charge-transfer (CT) excitation, which measures the unoccupied part of the O $2p$ partial density of states (PDOS). Because the energy position of the non-bonding O $2p$ band is located near the top of the O $2p$ band, we assumed it located ~2 eV below the Fermi level ($E_F$), based on the occupied O $2p$ PDOS deduced from fluorescence spectra as discussed below. Thus one finds the $e_g$ band 4-6 eV above $E_F$, the $j_{eff} = \frac{1}{2}$ band ~1 eV above $E_F$, the empty and occupied parts of the $j_{eff} = \frac{3}{2}$ band just above and below $E_F$, respectively.

The occupied part of the O $2p$ PDOS can be measured by the fluorescence component of the O $K$-edge RIXS. We have taken spectrum k in Fig. 2(d) as the representative fluorescence spectrum. The $E_F$ position for the fluorescence spectrum has been fixed under the assumption that $E_{\rm em}$ = 528.2 eV is the excitation threshold of the O $K$-edge XAS [photon energy c in Fig. 2(a)]. The combined occupied and unoccupied parts of the O $2p$ PDOS thus derived are plotted in Fig. 3(a).

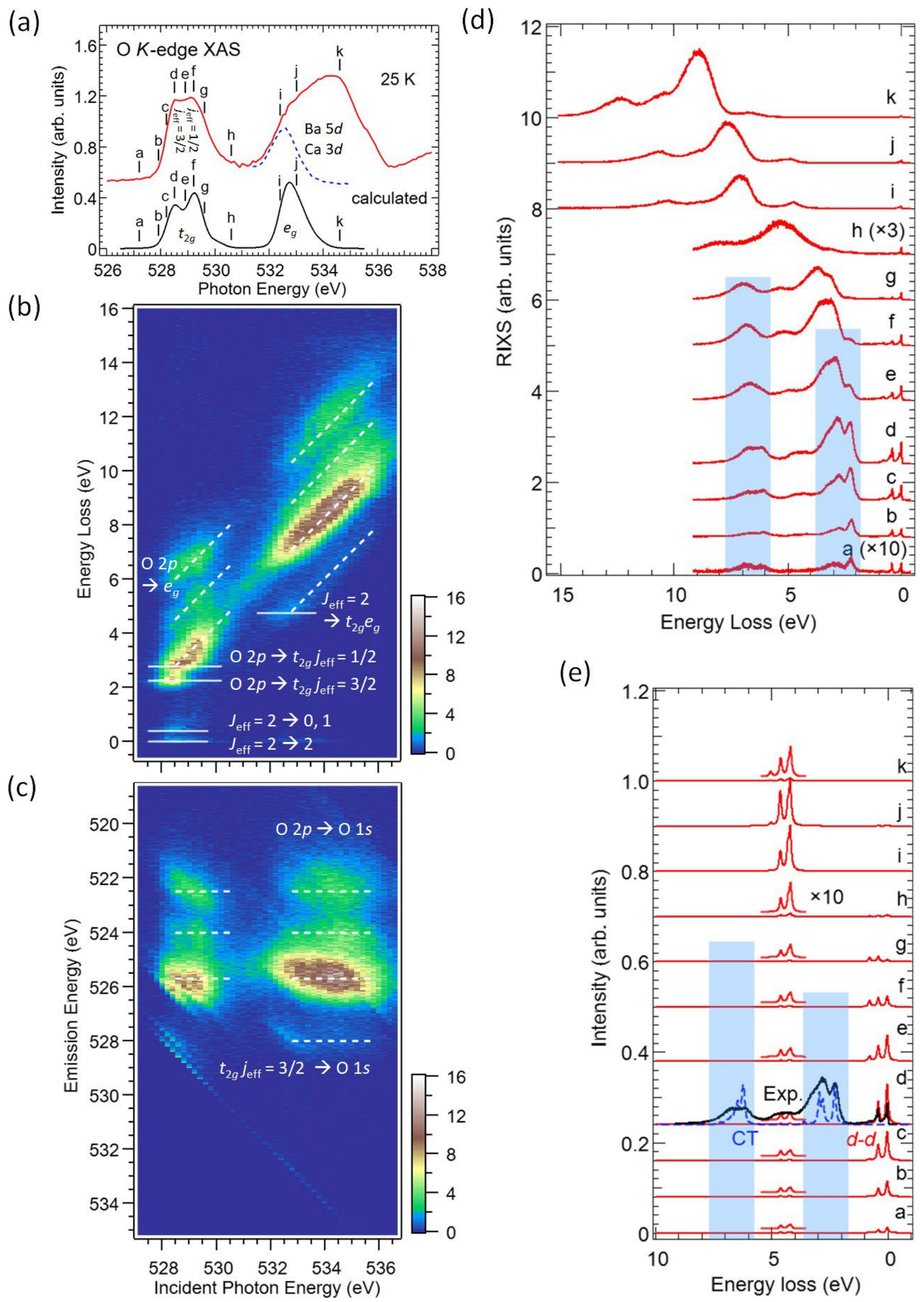

FIG. 2. X-ray absorption spectroscopy (XAS) and resonant inelastic X-ray scattering (RIXS) spectra of $Ba_2CaOsO_6$ at the oxygen $K$ edge recorded at 25 K. (a) XAS spectra. Top: XAS spectrum measured using the total fluorescence-yield method. The broad absorption band at 532-536 eV consists of transitions to the empty $e_g$ states (blue dashed curve) as well as to the Ba $5d$- and Ca $3d$-derived conduction-band states. The blue dashed curve is the partial fluorescence-yield spectrum measured at $E_{\mathrm{em}}$ = 528 eV. Bottom: Spectrum calculated using ligand-field (LF) multiplet theory. See Methods. (b) & (c) Colored intensity maps of scattered X-rays: (b) as a function of incident X-ray energy $E_{\mathrm{in}}$ and energy loss $E_{\mathrm{loss}}$. Solid lines mark the positions of constant $E_{\mathrm{loss}}$. Dashed lines indicate those of constant $E_{\mathrm{em}}$, (c) Scattered X-ray intensity map plotted against $E_{\mathrm{in}}$ and emission energy $E_{\mathrm{em}}$. Dashed lines indicate constant $E_{\mathrm{em}}$'s, corresponding to fluorescence from occupied states to the O $1s$ core level. (d) RIXS spectra measured for various $E_{\mathrm{in}}$'s indicated by vertical bars at the top spectrum of (a). The shaded parts mark transitions from O $2p \rightarrow$ Os $5d$ charge-transfer (CT) excitation. (e) LF multiplet calculation to simulate the RIXS spectra. The red curves show spectra arising from $d$-$d$ excitation for a series of $E_{\mathrm{in}}$'s indicated by vertical bars in the calculated O $K$-edge XAS spectrum shown at the bottom of (a). The dashed blue curves show O $2p \rightarrow$ Os $5d$ CT excitation simulated by $5d^2 \rightarrow 5d^3$ multiplet calculation. The black curve is the measured RIXS spectrum for $E_{\mathrm{in}}$ = 528.5 eV [spectrum d in panel (d)].

The obtained O $2p$ PDOS is compared with the DOS calculated by DFT (see Methods) in Fig. 3(b). One can see good one-to-one correspondence between the experimental and calculated structures in the O $2p$ PDOS. In particular, the assumed non-bonding O $2p$-band position well agrees with the peak position of the calculated non-bonding O $2p$ band. Nevertheless, the conclusion of the present paper will not altered by the magnitude of the band gap unless it collapses and the system becomes metallic.

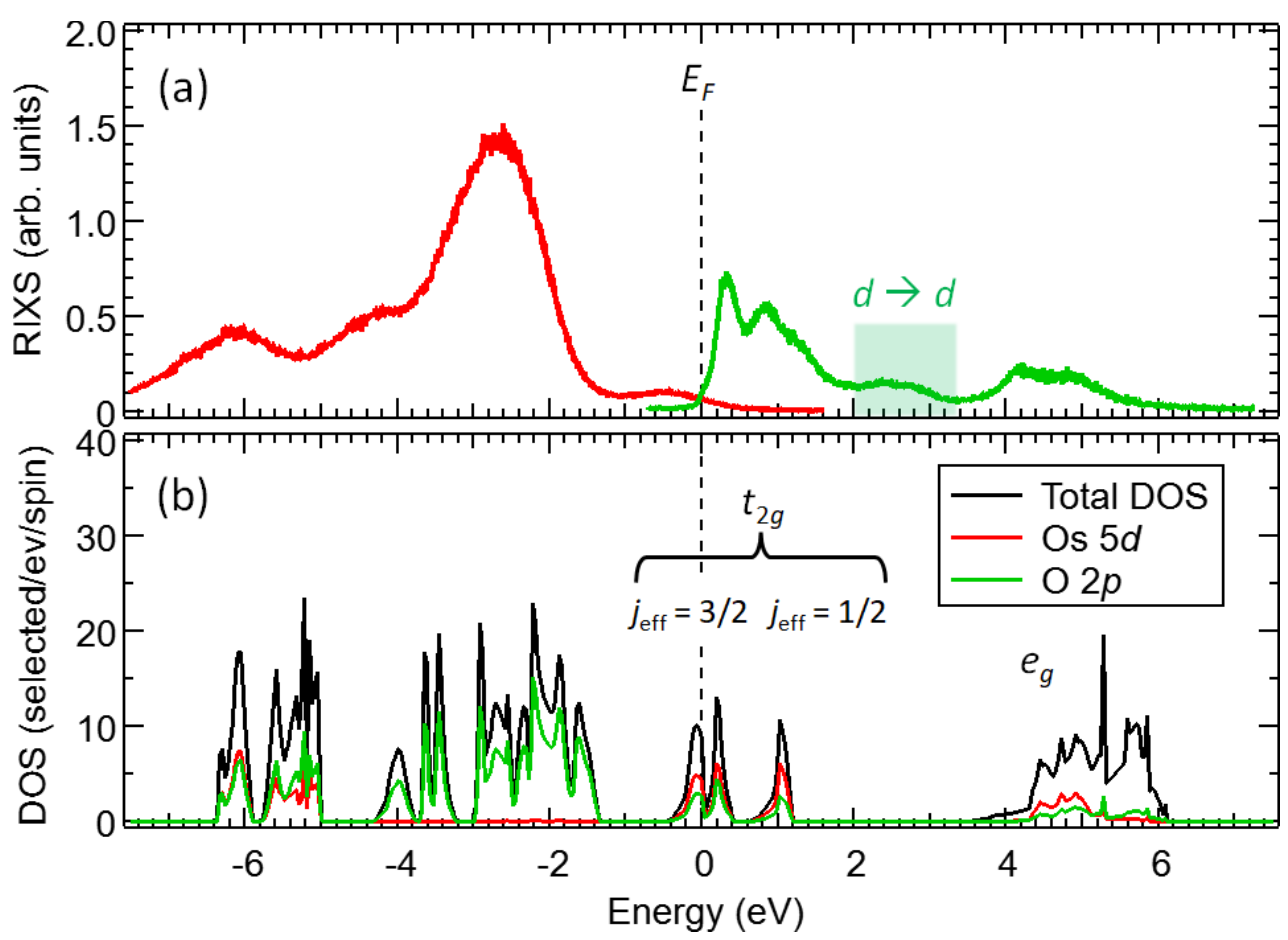

FIG. 3. O $2p$ partial density of states (PDOS) derived from experiment and theory. (a) Experimental O $2p$ PDOS. The empty part is derived from the O $2p \rightarrow$ Os $5d$ CT excitation in the RIXS spectrum [$E_{\rm in}$ = 528.2 eV, c in Fig. 2(d)], and the occupied part from the fluorescence component of the RIXS spectrum [$E_{\rm in}$ = 534.6 eV, k in Fig. 2(d)]. The shaded part marked by "$d \rightarrow d$" arises from $t_{2g}^2 \rightarrow t_{2g}e_g$ transition, and is unrelated to the O $2p$ PDOS. (b) PDOS of the nonmagnetic state obtained by the GGA+$U$+SOC calculation with $U - J_H$ = 2.5 eV.

### Low-energy multiplet and phonon satellites

In order to examine the effect of electron-phonon coupling and possible low-symmetry crystal field, an enlarged plot of the RIXS spectra in the low-energy region is shown in Fig. 4(a). The "elastic" peak at $E_{\rm loss}$ = 0 eV is a superposition of the genuine elastic scattering (which should be very weak for the present $\pi$ scattering geometry) and low-energy (0–20 meV) elastic and quasi-elastic scattering between the nearly degenerate five components of the $J_{\rm eff}$ = 2 ground state [see Fig. 1(b)]. The sharp peak at $E_{\rm loss} \sim$ 0.4 eV is due to excitation from the $J_{\rm eff}$ = 2 ground state to the $J_{\rm eff}$ = 0 and 1 excited states. The latter excitation is also observed by Raman scattering as described below. The weak peak at $E_{\rm loss} \sim$ 0.8 eV is an excited state of the $t_2^2$ multiplet having the quantum number $J_{\rm eff}$ = 2 [Fig. 1(b)]. Unfortunately, the residual cubic splitting $\Delta_{\rm c}$ of the $J_{\rm eff}$ = 2 state [Fig. 1(b)] is too small to be resolved in the RIXS spectra. Each of the quasi-elastic, $E_{\rm loss} \sim$ 0.4 eV, and $\sim$ 0.8 eV peaks are accompanied by sub-peaks with ~80 meV intervals on the higher-energy side. These sub-peaks are attributed to phonon replicas, as described below.

### Analyses using ligand-field multiplet theory

The magnitude of the SOC of the Os $5d$ states can be estimated from the O $K$-edge XAS [Fig. 2(a)]. X-ray absorption into the empty $t_{2g}$ state observed at 528–530 eV is split into double peaks separated by $\frac{3}{2}\zeta \sim$ 0.7 eV. Note that for the effective angular momentum operator $\mathbf{l}_{\rm eff}$, the SOC constant $\zeta'$, defined through the SOC energy $\zeta'\mathbf{l}_{\rm eff}\cdot\mathbf{s}$, is given by $\zeta' \equiv -\zeta$ because $\mathbf{l}_{\rm eff} \equiv -\mathbf{l}$ for the $t_{2g}$ electrons [26], where $\mathbf{l}$ is the bare angular momentum operator. The cubic ligand-field (LF) splitting $\Delta_{\rm LF}$ of the Os $5d$ level into $t_{2g}$ and $e_g$ [Fig. 1(a)] can also be estimated from the O $K$-edge XAS. From the broad absorption feature at 532–536 eV, the $e_g$ component could be isolated by monitoring the RIXS intensity of the $J_{\rm eff}$ = 2 ($t_{2g}{}^2$) $\rightarrow$ $t_{2g}e_g$ energy-loss feature at $E_{\rm loss}$ ~5 eV [marked by a solid line in Fig. 2(b)] as a function of $E_{\rm in}$: Thus obtained intensity plotted by the blue dashed curve at the bottom of Fig. 2(a) gives the $e_g$ component, allowing us to obtain $\Delta_{\rm LF} \simeq$ 4 eV.

To interpret the RIXS spectra quantitatively, LF multiplet calculations were performed and their results are shown in Fig. 2(e). (For details, see Methods.) The calculated $d^2$ multiplet (red curves) reproduces the observed loss peaks at $E_{\rm loss}$ ~0.4 and 0.8 eV ($t_2^2$ part), and those at $E_{\rm loss}$ ~4 eV ($t_{2g}e_g$ part). The RIXS spectra in the blue shaded parts, $E_{\rm loss} \simeq$ 2–4 and 6–7 eV, cannot be reproduced by the $d^2$ multiplet. We attribute these features to O $2p \rightarrow$ empty Os $5d$ CT excitation, and simulate the CT excitation spectrum by $d^2 \rightarrow d^3$ multiplet calculation of inverse-photoemission leaving a hole in the non-bonding O $2p$ band. By assuming that the non-bonding O $2p$ band is located $\sim$ 2 eV below $E_{\rm F}$, as indicated by the DFT calculation, and by ignoring the $p$-band width, we could reproduce the CT spectrum from the $J_{\rm eff}$ = 2 ground state and plotted it by dashed blue curves in Fig. 2(e). The $E_{\rm loss} \simeq$ 2–4 eV region arises from O $2p \rightarrow t_{2g}$ excitation and the $E_{\rm loss} \simeq$ 6–7 eV region arises from O $2p \rightarrow e_g$ CT excitation.

Note that there are no features in the RIXS spectra that indicate the lowering of the cubic symmetry: If the LF symmetry were lower than the cubic one, the $J_{\rm eff}$ = 2 ground state, which is split into the doublet $E_g$ and the triplet $T_{2g}$ [Fig. 1(b)], would be further split into multiple states as shown in Fig. 4(b) for cubic ($O_h$)-to-tetragonal ($D_{4h}$) symmetry lowering, and the low-energy part of the RIXS spectra would be significantly different from the ex-perimental ones in Fig. 4(a). If the tetragonal distortion were comparable to that of the $5d^1$ double perovskites $Ba_2NaOsO_6$ [22] and $Ba_2MgReO_6$ [23], which show peak at $E_{\mathrm{loss}}$ ~0.1 eV in the O $K$-edge RIXS spectra, the tetragonal crystal field $6Cp$, which is equal to the tetragonal splitting of the $t_{2g}$ level [25] should be as large as ±0.15 eV, and the quasi-elastic $J_{\mathrm{eff}} = 2 \rightarrow J_{\mathrm{eff}} = 2$ peak would be split into a few peaks over an energy range of ~0.1 eV in addition to the phonon satellites. The absence of temperature dependence in the spectral line shapes across the magnetic transition at $T^* \sim 50$ K suggests that the magnetic transition does not involve any appreciable structural change. Furthermore, there are no spectral features that can be attributed to magnons nor bi-magnons, consistent with the absence of spin order in $Ba_2CaOsO_6$.

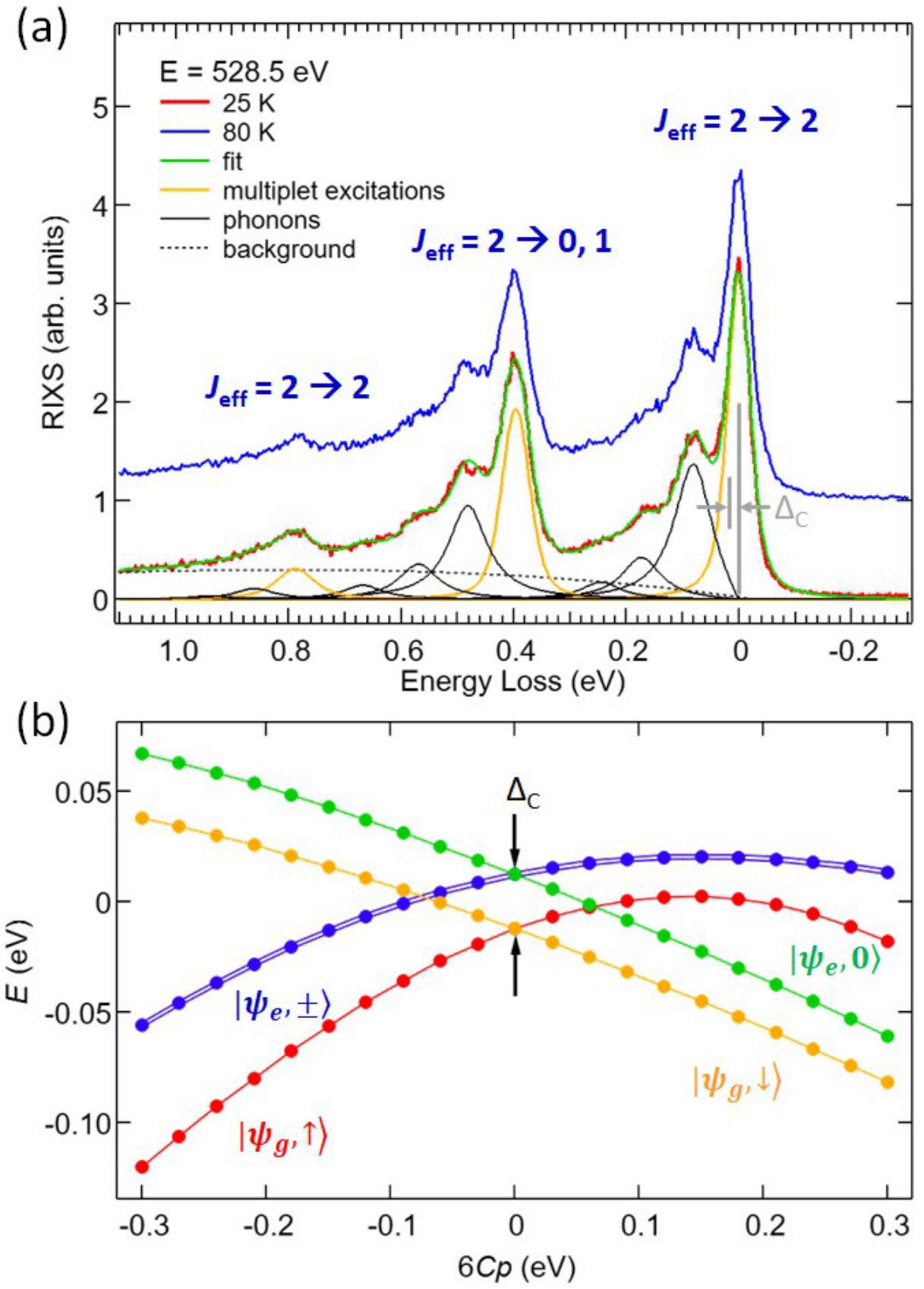

FIG. 4. Low-energy excitations measured by O $K$-edge RIXS. $\Delta_c$ is the 'residual' cubic splitting of the $J_{\mathrm{eff}} = 2$ ground state defined in Fig. 1(b). (a) RIXS spectra in the low energy-loss region at different temperatures across the magnetic transition at $T^* \sim 50$ K for the incident photon energy of 528.5 eV. The $d$-$d$ excitation shown in Fig. 1(b) is seen. The energy-loss peaks as well as the quasi-elastic peaks are accompanied by multiple phonon satellites. The splitting $\Delta_c \sim 20$ meV is not resolved in the spectra. Details of the line-shape analysis are given in Supplementary Note 1 with Supplementary Figure 1 and Supplementary Table 1. (b) Energies of low-lying excited states of the $Os^{6+}$ ($5d^2$) ion as functions of the low-symmetry $D_{4h}$ LF parameter $6Cp$ [25], the splitting of the $t_{2g}$ level. The $O_h$ LF parameter $\Delta_{\mathrm{LF}}$ is set to 4.1 eV. $|6Cp| \simeq 0.15$ reproduces the low-energy excitation in the O-$K$ RIXS spectra of the $5d^1$ double perovskites [22, 23].

The octupolar nature of the ground state of the $Os^{6+}$ ion in the $O_h$ field can be demonstrated by LF multiplet calculation with a weak magnetic field in the (1,1,1) direction as a time-reversal symmetry-breaking perturbation that splits the non-Kramers $E_g$ doublet into $|\psi_{g,+}\rangle$ and $|\psi_{g,-}\rangle$. For $B = 10$ T, we obtained octupolar moment $T_{xyz} \equiv \langle \overline{J^x J^y J^z} \rangle \simeq 1.2$ with a tiny dipole magnetic moment of $\sim 8 \times 10^{-3} \mu_{\mathrm{B}}$ induced along the (1,1,1) direction on top of the octupolar ground state. For a smaller field of $B = 1$ T, the induced dipolar moment was as small as $\sim 7 \times 10^{-4} \mu_{\mathrm{B}}$. Generally, for a small field $B \ll \Delta_c / \mu_{\mathrm{B}}$, the induced magnetic dipole moment along the direction of $B$ ($\sim \mu_{\mathrm{B}} B / \Delta_c$) is proportional to $B$. While ordering with the finite magnetic octupolar or electric quadrupolar moment can occur within the $E_g$ ground state, the appearance of a finite magnetic dipolar moment requires hybridization between the $E_g$ and $T_{2g}$ states. Thus, to obtain a stable magnetic dipole order by the exchange interaction, a molecular field of $\mu_{\mathrm{B}} B > \Delta_c$ at least is required.

## Phonon Raman scattering

To further confirm the absence of low-symmetry crystal field, we employed Raman scattering spectroscopy, a sensitive probe of lattice symmetry. Figure 5(a) shows one-phonon Raman spectra of a $Ba_2CaOsO_6$ polycrystal taken at 80 K with two polarization geometries (∥ and ⊥; for technical details, see Methods). There must be phonons of $A_{1g} + E_g + T_{1g} + 2T_{2g} + 5T_{1u} + T_{2u}$ symmetries at the Brillouin-zone center in case of cubic $Fm\bar{3}m$ structure, out of which four ($A_{1g}$, $E_g$, and $2T_{2g}$) phonons are Raman-active.

To obtain information about the symmetry of the observed excitation, polarization measurements were performed in two geometries - with parallel (∥) and with mutually perpendicular (⊥) polarizations of the incident and scattered light. We utilized the rules that in isotropic or cubic systems the depolarization ratio $\rho = I_{\perp} / I_{\parallel}$ does not exceed 0.75 for totally symmetric modes, while it is close to 0.75 for non-totally symmetric ones [27]. One can see that the line at ~796.5 $cm^{-1}$ obviously dominates in the ∥ spectrum and can be assigned to the $A_{1g}$ mode, while the phonons at frequencies 102.5, 375, and 495 $cm^{-1}$ are observed in both polarized ∥ and depolarized ⊥ spectra and are assigned to $T_{2g}$, $T_{2g}$, and $E_g$ modes, respectively, with the help of the non-magnetic DFT calculation of phonon modes as described in Supplementary Note 2 with Supplementary Figure 2. The broad peak at 720 $cm^{-1}$ has a fairly low depolarization ratio (~0.2), which suggests its $A_{1g}$ symmetry. While its origin is not clear, the symmetry lowering cannot split the $A_{1g}$ mode without the increase of the unit cell. (Note also that the second intensive $T_{2g}$ mode remains unsplit.) The appearance of this low intensity $A_{1g}$ peak can be, e.g., due to imperfections of the crystal structure (e.g. there are indications of anti-phase boundary defects with disorder at the $B$ sites and other types of defects [28–31]) or a two-phonon repetition of 375 cm$^{-1}$ vibration. Thus, our Raman experiments do not detect any direct evidence of the symmetry lowering in the cubic $Ba_2CaOsO_6$.

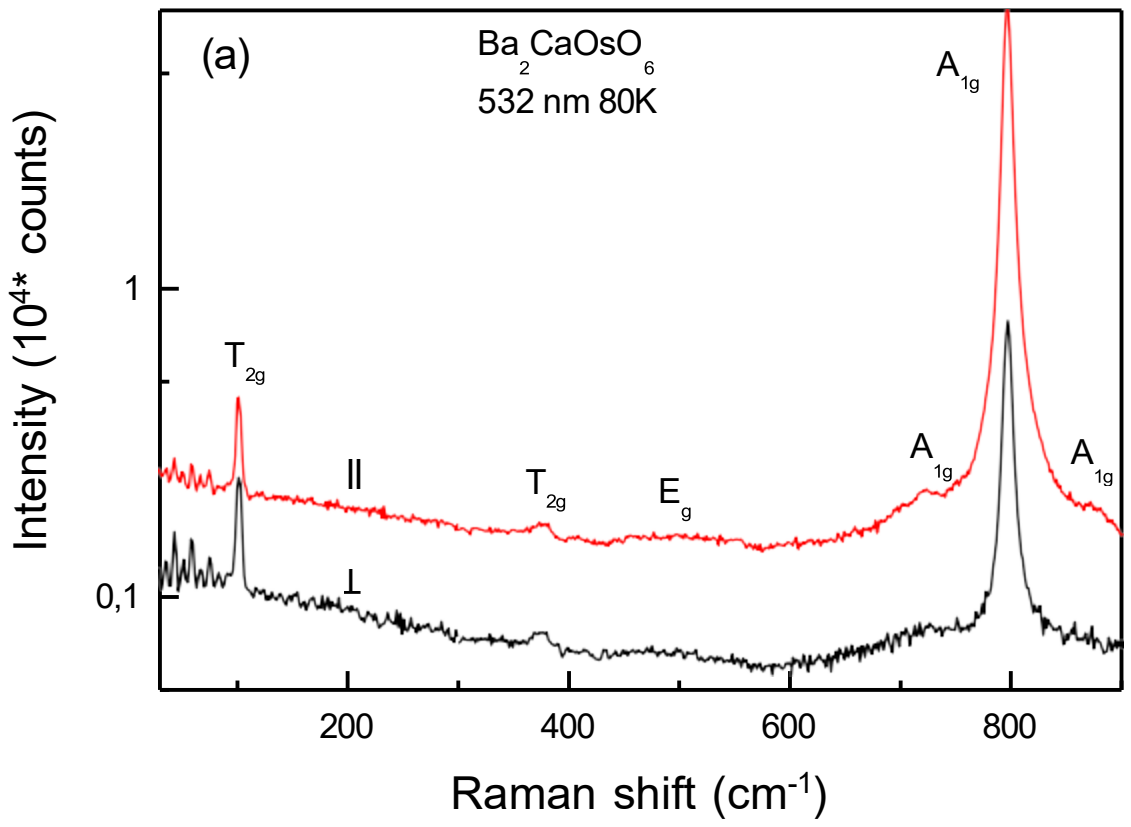

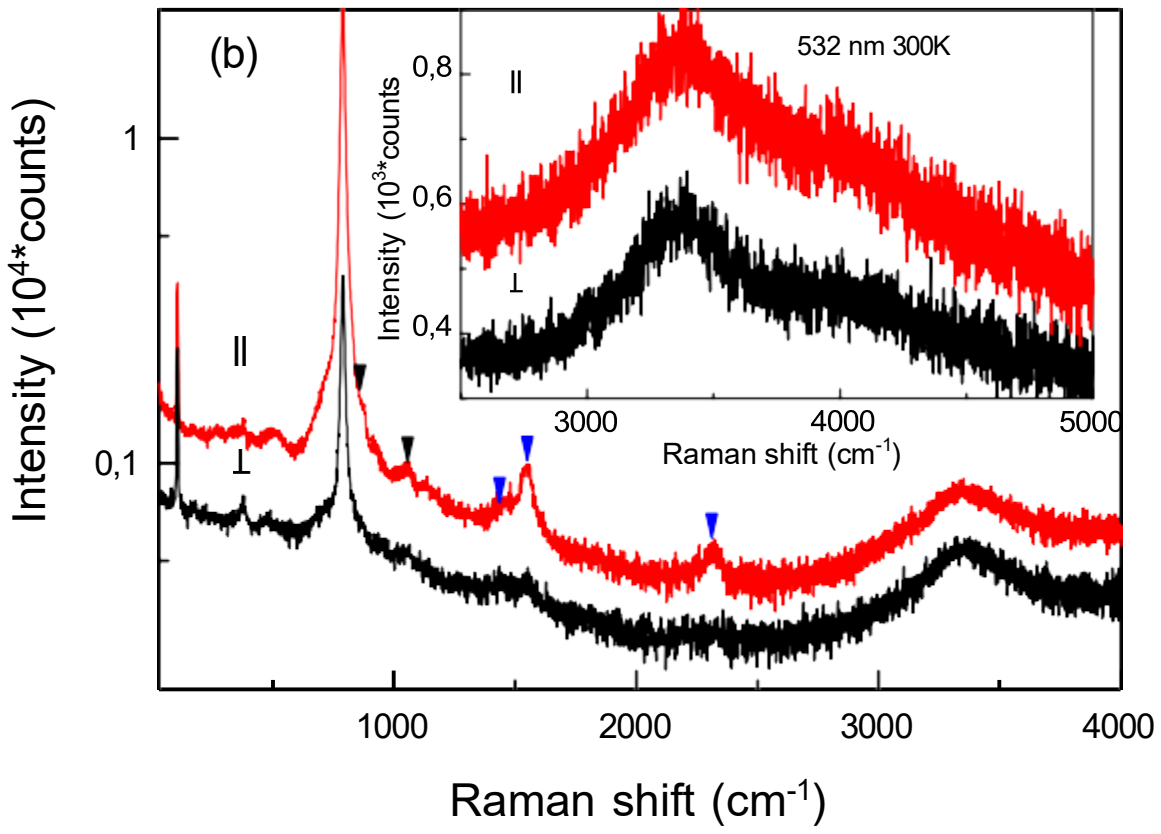

FIG. 5. Raman spectra in two scattering symmetries: ∥ (red) and ⊥ (black) are presented. (a) Frequency range with a one-phonon excitations. (b) Extended frequency range where the high-order phonon processes (blue arrows show features originating from the $A_{1g}$ branch, black - another weak two-phonon features) and electronic excitation are seen. Latest are seen in the range from 3000 to 5000 cm$^{-1}$, as shown in the inset, most probably due to the $J_{\rm eff}=2$ ground state to the $J_{\rm eff}=1,0$ excited states seen by RIXS (Fig. 4).

To examine the possible lattice distortions across $T^* \sim$ 50 K, we measured spectra at 10 K with better resolution and found only minor changes in the spectrum, such as further hardening and narrowing of the $A_{1g}$ mode at ~796.5 cm$^{-1}$. Interestingly, the frequency and unique line width of the low-frequency $T_{2g}$ mode did not change in the whole temperature range of 10-300 K.

The higher frequency range from 800 to 3000 cm$^{-1}$ presented in Fig. 5(b) shows several peaks dominating in the ∥ geometry and hence of the $A_{1g}$ symmetry. The weak peaks at 870 and 1055 cm$^{-1}$ indicated by black arrows are probably two-phonon features. The peaks at 1450, 1540 and 2310 cm$^{-1}$ shown by blue arrows can be associated with double and triple phonon scattering from the $A_{1g}$ phonon branch. At higher frequencies (>3000 cm$^{-1}$), two broad peaks are observed. They are clearly seen in both polarization geometries (∥ and ⊥), as well as upon excitation by both used laser lines (532 and 633 nm, see Methods), which indicates Raman scattering by electron excitations of $T_{2g}$ or $E_g$ symmetry. Their energies agree well with the $J_{\rm eff}=2 \rightarrow J_{\rm eff}=0, 1$ RIXS peaks (Fig. 4).

Typically, in the case of $5d^2$ double perovskites, the effect of tetragonal distortions on the ground state is considered due to their stronger coupling with electronic structure [32]. Interestingly, Rayyan *et al.* [33] included the trigonal distortions in their analysis and demonstrated that, while these distortions are unable to split the $E_g$ doublet due to symmetry constraints, they can lift the degeneracy of the higher-lying $T_{2g}$ triplet. Under trigonal distortions, some of these $T_{2g}$ states go to a lower energy. However, as shown in [33] there is a rather wide range of trigonal distortion under which the non-Kramers $E_g$ doublet remains the ground state. This doublet may, therefore, survive as the ground state under a small lattice distortion. It is also possible that distortion is substantially reduced due to dynamical Jahn-Teller effect. Whether the $E_g$ doublet hosts an electric quadrupole or a magnetic octupole or both depends on the type of broken symmetry (spatial symmetry or time-reversal symmetry) and the strength of the mean field from neighboring Os ions.

### Electron-phonon coupling effects on RIXS

In the low-energy RIXS spectra shown in Fig. 4, the quasi-elastic peak and the peak at $E_{\rm loss} \simeq 0.4$ eV are accompanied by sub-peaks separated by ~80, 160, and 240 meV with decreasing intensities. We attribute the sub-peaks to phonon replicas created by the simultaneous excitation of optical phonons. The one-phonon energy of ~80 meV is somewhat lower but in a similar range as the Raman $A_{1g}$ mode energies 720 cm$^{-1}$ = 88 meV and 796.5 cm$^{-1}$ = 99 meV. The replica energies are close to those obseved in the RIXS spectra of $Ba_2NaOsO_6$ [22]. From the replica intensities, the dimensionless electron-phonon coupling constant is estimated to be $M/\omega_0 \gtrsim 1$, where $M$ is the average electron-phonon coupling matrix element and $\omega_0$ is the phonon energy [34]. In spite of the moderately strong coupling, Jahn-Teller distortion is suppressed in $Ba_2CaOsO_6$ due to the strong SOC and dynamical Jahn-Teller effect, suggesting that Os-Os exchange interaction is strong enough to stabilize the magnetic octupole over the electric quadrupole. Here, it should be noted that sub-peaks similar to the phonon replicas may appear in the RIXS spectra if dynamical Jahn-Teller effect [35] exists, as reported for the $5d^1$ sys-tem $Ba_2CaReO_6$ [18, 19]. Whether such an effect also exists in $5d^2$ systems or not is an interesting question to be pursued in future studies. Considering the different time scales of RIXS ( on the order of 0.01 fs) and Raman scattering (> 10 fs), it is possible that dynamical Jahn-Teller effect was seen in RIXS as the "phonon replicas" but not in Raman scattering.

TABLE I. Parameter values for the Os $5d$ electrons hybridized with O $2p$ orbitals in $Ba_2CaOsO_6$ derived in the present XAS and RIXS spectra.

| Parameter | Symbol | Value (eV) |
| --- | --- | --- |
| ligand-field splitting | $\Delta_{LF}$ | 4.1 |
| Spin-orbit coupling for the $5d$ shell | $\zeta$ | 0.47 |
| Spin-orbit coupling for the $t_{2g}$ shell | $\zeta'$ | -0.47 |
| Hund's coupling for the $t_{2g}$ shell | $J_H$ | 0.27 |

## III. DISCUSSION

We have investigated the electronic structure of $Ba_2CaOsO_6$ by XAS, RIXS, and Raman scattering experiment as well as DFT calculation, focusing on extracting reliable parameters characterizing the systems, as summarized in Table I, and on the confirmation of the local cubic symmetry of the Os ions that favors the octupolar state as the origin of its 'hidden order'. We have also confirmed the octupolar nature of the non-Kramers $E_g$ doublet ground state ($|S_z| = 0$ and $|L_z| = 0$ under an infinitesimally small magnetic field) by LF multiplet calculation.

Owing to the hybridization between the O $2p$ and Os $5d$ orbitals, electronic excitation within the $5d^2$ multiplet and charge-transfer excitation from the occupied O $2p$ to the empty Os $5d$ states could be identified by the O $K$-edge RIXS. From comparison of the XAS and RIXS line shapes with the LF multiplet calculation, the absence of splitting of low-energy RIXS peaks as well as the lack of additional lines in Raman scattering spectra we conclude that no crystal field lower than the cubic one can be identified, consistent with the small (~20 meV) residual cubic splitting of the $J_{\rm eff} = 2$ ground state. The present results obtained by different types of X-ray and optical spectroscopy, which are typically very sensitive to a local environment of transition metals, substantially strengthen previous findings, in particular diffraction data demonstrating the absence of non-cubic distortions [14].

There are two possible mechanisms working hand-in-hand in suppressing the Jahn-Teller distortion expected for the $Os^{6+}$ ion with the $d^2$ configuration. In both mechanisms, the strong SOC is involved. One is an on-site effect related to the stabilization of electrons not at cubic harmonics as the crystal field (i.e. Jahn-Teller effect) would prefer, but rather on entangled spin-orbitals [4, 36]. Our RIXS measurements clearly resolved phonon replicas of the $J_{\rm eff} = 2 \rightarrow 1, 0, 2$ excitation peaks. This allowed us to estimate the electron-phonon coupling strength, which turns out to be moderately strong, $M/\omega_0 \gtrsim 1$ and, therefore, may not be sufficiently strong to recover the Jahn-Teller distortion but might induce dynamical Jahn-Teller effect. On the other hand, there is also inter-site effect – the energy gain due to exchange interaction between the octupoles, which are formed by SOC. Further spectroscopic and theoretical studies are necessary to identify the octupolar order and its microscopic origin.

## METHODS

### Materials preparation

Polycrystalline $Ba_2CaOsO_6$ was synthesized through a solid-state reaction using fine powders of $BaO_2$ (99% purity, Kojundo Chemical Laboratory Co., Ltd.), $CaO_2$ (prepared in the laboratory [37]), and Os (99.95% pu- rity, Nanjing Dongrui Platinum Co. Ltd.) in a ratio of 2:1:1.1. Approximately 200 mg of the mixed materi- als were placed into an alumina crucible. The mixture was then heated in air to 1000°C for 7 hours, followed by a 1-hour dwell time, and subsequent cooling to room temperature over a span of 7 hours. After re-mixing and pressing, the sample was annealed at 1000°C for 24 hours. The resulting product is a grey sintered pellet, possessing sufficient solidity to be manipulated with tweezers. Powder X-ray diffraction analysis was performed using Cu $K\alpha$ radiation within the $5^\circ \leq 2\theta \leq 65^\circ$ range at 293 K. The measurements were conducted with a MiniFlex600 diffractometer (Rigaku, Tokyo, Japan). The acquired data, shown in Supplementary Figure 3, exhibited good agreement with simulations based on the crystallographic data of $Ba_2CaOsO_6$ [12], confirming the single-phase nature of the product.

### Resonant inelastic X-ray scattering

All resonant inelastic X-ray scattering (RIXS) and X-ray absorption spectroscopy (XAS) measurements at the O $K$ edge were performed using the AGM-AGS spectrometer of beamline 41A at Taiwan Photon Source of National Synchrotron Radiation Research Center (NSRRC) [38]. This beamline is based on the energy compensation principle of grating dispersion [39]. The energy bandwidth of incident X-ray was 0.2 eV (0.1 eV for XAS measurement) while keeping the total energy resolution of RIXS as 30 meV at the incident photon energy of 528.5 eV. The sample surface was cleaned by scraping with a diamond file in the Ar glove box before the measurement and was transferred into the measurement chamber without exposure to the air. The base pressure of the measurement chamber was $\leq 1 \times 10^{-8}$ Torr. The sample was cooled down to 25 K with liquid helium during the measurements. Both RIXS and XAS measurements were carried out using linear horizontally ($\pi$) polarized X-rays. The XAS spectra were measured with a normal-incident X-ray in the total fluorescence yield mode. For the RIXS measurement, the incidence angle was fixed at 20$^\circ$, and the scattering angle was fixed at 90$^\circ$. The combination of the $\pi$-polarized X-rays and the 90$^\circ$ scattering angle makes the RIXS signals purely magnetic. The same geometry also allowed us to reduce the elastic peak and to study low-energy excitation effectively.

### Ligand-field multiplet calculation

Ligand-field multiplet (LF) calculations were performed by using the XTLS 8.5 package [40]. In the calculation of the O $K$-edge RIXS spectra, we assumed that the excited states of the $5d^2$ multiplet can be reached by O $K$-edge RIXS through the strong Os $5d$-O $2p$ hybridization and could be simulated by the calculation of Os $L_{2,3}$-edge RIXS by setting the $2p$-$5d$ Slater integrals and the Os $2p$ core-level SOC to zero. While this simulation would give the energy positions of RIXS features correctly, it would not give correct intensities because relevant transition-matrix elements are not used. The O $K$-edge XAS [Fig. 2(a)] was also simulated by the Os $L_{2,3}$-edge XAS in the same manner.

In general, the Slater integrals $F$'s and $G$'s (anisotropy of Coulomb interaction) and the SOC coupling constant $\zeta$ in solids are smaller than those of isolated atoms, because the wavefunctions are more spatially extended due to hybridization. In order to model this effect, the atomic Slater integrals and $\zeta$, deduced from Hartree-Fock calculations [41, 42], were multiplied by constant factors $R_{\mathrm{Slater}}$ and $R_{\mathrm{SOC}}$ ($0 \leq R_{\mathrm{Slater}} < 1$, $0 \leq R_{\mathrm{SOC}} < 1$), respectively. These factors $R_{\mathrm{Slater}}$ and $R_{\mathrm{SOC}}$ and the cubic LF splitting $\Delta_{\mathrm{LF}}$ and were treated as adjustable parameters. For O $K$-edge RIXS, $\zeta = 0.50$ eV and $\Delta_{\mathrm{LF}} = 4.1$ eV were used, and the Slater integrals between the Os $5d$ orbitals were reduced to 35% of the atomic Hartree-Fock values. Hund's coupling $J_{\mathrm{H}}$ between two $d$ electrons (Table I) is related to Slater integrals through $J_{\mathrm{H}} = \frac{3}{49}F^2 + \frac{20}{441}F^4$ [43]. The value $J_{\mathrm{H}} = 0.27$ eV in the table is smaller than $J_{\mathrm{H}} = 0.5$ eV used for the DFT+$U$ +SOC calculation because the former is for the Os $5d$-O $2p$ anti-bonding orbitals while the latter for the Os $5d$ atomic orbitals. The reduction of $J_{\mathrm{H}}$ from 0.5 eV to 0.27 eV suggests that the atomic orbitals consisting of the antibonding $t_{2g}$ band have the weight Os $5d$ : O $2p$ ~ 70%: 30%.

In the calculation of the RIXS spectra, the same geometry as the experiment was adopted: The incident and scattered X-rays were set parallel to the cubic [001] and [100] directions, respectively. Taking the [001], [100], and [010] directions as the $z$, $x$, $y$ axes, respectively, the linear polarizations of the incident and scattered X-rays were set to be ($x$, $y$) and ($x$, $z$), and the spectra for these two polarization sets were averaged.

The calculated spectra were broadened by a Voigt function, which is the convolution of a Lorentz function and a Gauss function. The widths (half width at half maximum, HWHM) of the Lorentz functions were determined from the natural lifetime of the core holes: 0.05 eV for the O $K$-edge RIXS [44]. The widths (standard deviation) of the Gauss functions were assumed to be 0.01 eV. The XAS and RIXS spectra were calculated for the five lowest states [the lowest $J_{\mathrm{eff}} = 2$ state in Fig. 1(b)] as the initial state and were summed up according to the Boltzmann distribution of the initial states.

### Raman spectroscopy

Raman measurements in the 10-300K range were performed in backscattering geometry from the polycrystalline sample using an RM1000 Renishaw microspectrometer equipped with a 532 nm solid-state laser and 633 helium–neon laser. Very low power (up to 1 mW) was used to avoid local heating of the sample. A pair of notch filters with a cut-off at 60 cm$^{-1}$ were used to suppress light from the 633 nm laser line. To reach as close to the zero frequency as possible, we used a set of three volume Bragg gratings (VBG) at 532 nm excitation to analyze the scattered light. The resolution of our Raman spectrometer was estimated to be 2–3 cm$^{-1}$.

The temperature dependence of the two narrow lines in the spectrum [Fig. 5 (a)] turned out to be opposite. The fully symmetric line softened from 797.5 to 788.5 cm$^{-1}$ with an increase in the temperature range from 10 to 300 K, and its width increased from 6.5 to 12 cm$^{-1}$, which can be explained by anharmonicity effects. In contrast to this behavior, the energy and width of the low-frequency $T_{2g}$ phonon line remains constant within the measurement error ($\omega$ ~102.5 cm$^{-1}$ and $\Gamma$ ~1.5 cm$^{-1}$) when heated from 10 to 300 K. Unfortunately, the temperature behavior of the two broad lines - $T_{2g}$ at 375 cm$^{-1}$ ($\Gamma \sim$ 20 cm$^{-1}$) and $E_g$ at 495 cm$^{-1}$ ($\Gamma \sim$ 60 cm$^{-1}$) - is difficult to study due to their weak intensity. However, despite the large width of the lines, we did not find any signs of their splitting.

### Density-functional-theory calculation

The generalized gradient approximation (GGA) in the form proposed by Perdew, Burke, and Ernzerhof [45] as realized in VASP code [46] was used for the density functional theory calculations. Phonon spectra shown in Supplementary Figure 2 were calculated by the frozen phonon method [47] with $5 \times 5 \times 5$ mesh of the Brillouin zone of the $2 \times 2 \times 2$ supercell in non-magnetic GGA. Planewave cut-off was set up to 500 eV. The structure was relaxed until convergence in energy of $10^{-6}$ eV in electronic subsystem and $10^{-5}$ eV in ionic one was achieved.

## DATA AVAILABILITY

All data generated or analysed during this study are available from the corresponding authors upon reasonable request.

## ACKNOWLEDGEMENTS

We are grateful to M. Haverkort, A. Paramekanti, C. Franchini, and A. Hariki for useful discussions. The soft X-ray measurements were conducted at beam line 41A of Taiwan Photon Source. This work was partly supported by the National Science and Technology Council of Taiwan under Grant Nos. 103-2112-M-213-008-MY3, 108-2923-M-548 213-001, and 113-2112-M-007-033 and by the Japan Society for the Promotion of Science under Grant Nos. JP20K14416 and JP22K03535. We thank the Ministry of Science and Higher Education of the Russian Federation for supporting Raman experiments and their interpretation through funding the Institute of Metal Physics, while theoretical calculations were performed with aid of the Russian Science Foundation (project RSF 23-42-00069). A.F. acknowledges the support from the Yushan Fellow Program under the Ministry of Education of Taiwan.

## AUTHOR CONTRIBUTIONS

A.F., S.V.S. and D.J.H. coordinated the project. J.O., H.Y.H., A.S., D.J.H. and C.T.C. developed the RIXS instruments and conducted the RIXS experiments. Y.S.P. performed Raman experiments. H.H. and K.Y. synthesized and characterized the sample. G.S. and A.T. performed multiplet calculations. J.O., D.J.H., S.V.S. and A.F. analyzed the data and wrote the paper with inputs from other authors.

## COMPETING INTERESTS

The authors declare that there are no competing interests.

Supplementary Information for

# Spin-orbit-entangled state of $Ba_2CaOsO_6$ studied by O *K*-edge resonant inelastic X-ray scattering and Raman spectroscopy

J. Okamoto, G. Shibata, Yu. S. Ponosov, H. Hayashi, K. Yamaura, H. Y. Huang, A. Singh, C. T. Chen, A. Tanaka, S. V. Streltsov, D. J. Huang, and A. Fujimori

**This SI file includes:**

Supplementary Notes 1 to 2

Supplementary Figures 1 to 3

Supplementary Table 1

Supplementary References 1 to 5

# Supplementary Note 1. Phonon sidebands in RIXS spectra

In the low-energy O $K$-edge RIXS spectrum of $Ba_2CaOsO_6$ in Fig. 4(a), subpeaks separated by an interval of $\sim$ 80-meV are observed to accompany the quasi-elastic peak and the peaks at $E_{\text{loss}} \simeq 0.4$ eV and 0.8 eV with intensities decreasing with energy. We attribute these sub-peaks to phonon replicas created by the simultaneous excitation of optical phonons. We investigated these multiplet excitation peaks and the phonons below energy loss of 1.5 eV referring to Živkovic´ *et al.*'s method [1]: Three multiplet excitation peaks and phonon side bands are fitted by Voigt functions and anti-Lorentzian functions, respectively. The background curve is a cubic function. Parameter values for the three multiplet excitations with Voigt functions are summarized in Supplementary Table 1: peak energy $\omega_0$, Gaussian FWHM $\Gamma_G$, integrated intensity, and the FWHM ratio of the Lorentzian function against the Gaussian function $\Gamma_L/\Gamma_G$.

For the phonon satellites, the relative peak energies and intensities of the phonon satellites are shown in Supplementary Figures 1(a) and (b). The FWHMs of the anti-Lorentzian functions are set to 85 meV. The relative peak energies of the phonon satellites $E_n^a$($a$ = 1st, 2nd, and 3rd multiplet peaks) are well approximated by linear functions of phonon number $n$, $E_n^a = nE_{\text{mode}}$ with the energy of the optical phonon mode $E_{\text{mode}}$ = 82±8 meV. From the intensity ratio of the phonon sidebands, the second harmonic intensities are $\sim$ 0.3 − 0.4 of the first harmonic ones. Then, one can estimate the parameter of the electron-phonon coupling strength $M/\omega_0 \gtrsim 1$ referring to [2].

# Supplementary Note 2. DFT calculations of phonon spectrum

In order to confirm the assignment of the phonon modes, we performed phonon calculations by the frozen phonons method [3] within the non-magnetic DFT. Os is a heavy transition-metal atom and, therefore, it is natural to include both strong electronic correlations ($U$) and the spin-orbit coupling (SOC) via DFT+$U$+SOC approach. However, this method tends to stabilize magnetic solutions and cannot directly simulate the non-Kramers many-electron $E_g$ states with zero projected total angular momentum $J^z_{eff} = 0$, which were proposed to be the ground state of $Ba_2CaOsO_6$ in case of cubic symmetry. Therefore, in Supplementary Figure 2 we present results of the non-magnetic phonon calculations [3]. The results of DFT+$U$ +SOC approach shall be discussed below.

Phonon frequencies at the Γ-point were calculated by density functional perturbation theory (DFPT) in the DFT+$U$ +SOC approach [4]. The same convergence criteria and parameter setup was used as in the frozen phonon calculations. On-site Hubbard repulsion parameter $U$ and Hund's intra-atomic exchange ($J_H$) were chosen to be $U = 3$ eV and $J_H = 0.5$ eV, which are close to what is used for Os ions in the literature. The crystal structure was taken from [5]. We tested several combinations of magnetic orders [ferromagnetic and antiferromagnetic (AFM) of A-type] and directions of the total momentum ([001], [110], [111]) and obtained that the lowest total energy corresponds to Immm structure with Os AFM-A (ferromagnetic planes) and magnetic moments ordered in the $ab$-plane close [110] direction. In the relaxed structure $OsO_6$ octahedra are slightly elongated: four short 1.930 A and two long 1.945 A Os-O bonds.

Results of phonon calculations are shown in Supplementary Figure 2. The experimental spectrum agrees well with the oversimplified non-magnetic DFT calculation (except for the $E_g$ mode). However, theoretical results for the cubic structure do not indicate any additional $A_{1g}$ phonons at 720 cm$^{-1}$ seen in the experiment.

Account of the spin-orbit coupling and Coulomb correlations by non-magnetic DFT+$U$+SOC ($U - J_H = 2.5$ eV) calculations of phonon frequencies was performed only at the Γ-point using DFPT. It improves the position of the $E_g$ band and yields $\omega_{A_{1g}} = 787$ cm$^{-1}$ (experiment: 796.5 cm$^{-1}$), $\omega_{E_g} = 525$ cm$^{-1}$ (experiment: 495 cm$^{-1}$), $\omega_T = 355$ cm$^{-1}$ (experiment: 375 cm$^{-1}$), and $\omega_{T_{2g}} = 105$ cm$^{-1}$ (experiment: 101.5 cm$^{-1}$). However, as explained above the undistorted cubic structure turns out unstable in GGA+U+SOC therefore there appear imaginary acoustic and low-frequency optical modes. The DFPT calculations demonstrate that the $E_g$ phonon mode splits, but the splitting does not exceed 30 cm$^{-1}$, so one cannot attribute two experimentally observed peaks at 495 and 720 cm$^{-1}$ to $E_g$ phonon split due to the Jahn-Teller effect. Nevertheless, experimental peak at 495 cm$^{-1}$ is extremely broad and therefore the splitting of this line can remain unnoticed. Unexpected appearance of the extra $A_{1g}$ mode at 720 cm$^{-1}$ can be due to a non-negligible disorder in $B$ sites. Indeed folding of the Brillouin zone, e.g. when $U \rightarrow \Gamma$, would lead to the appearance of an additional (weak) mode at ∼90 meV.

# Supplementary References

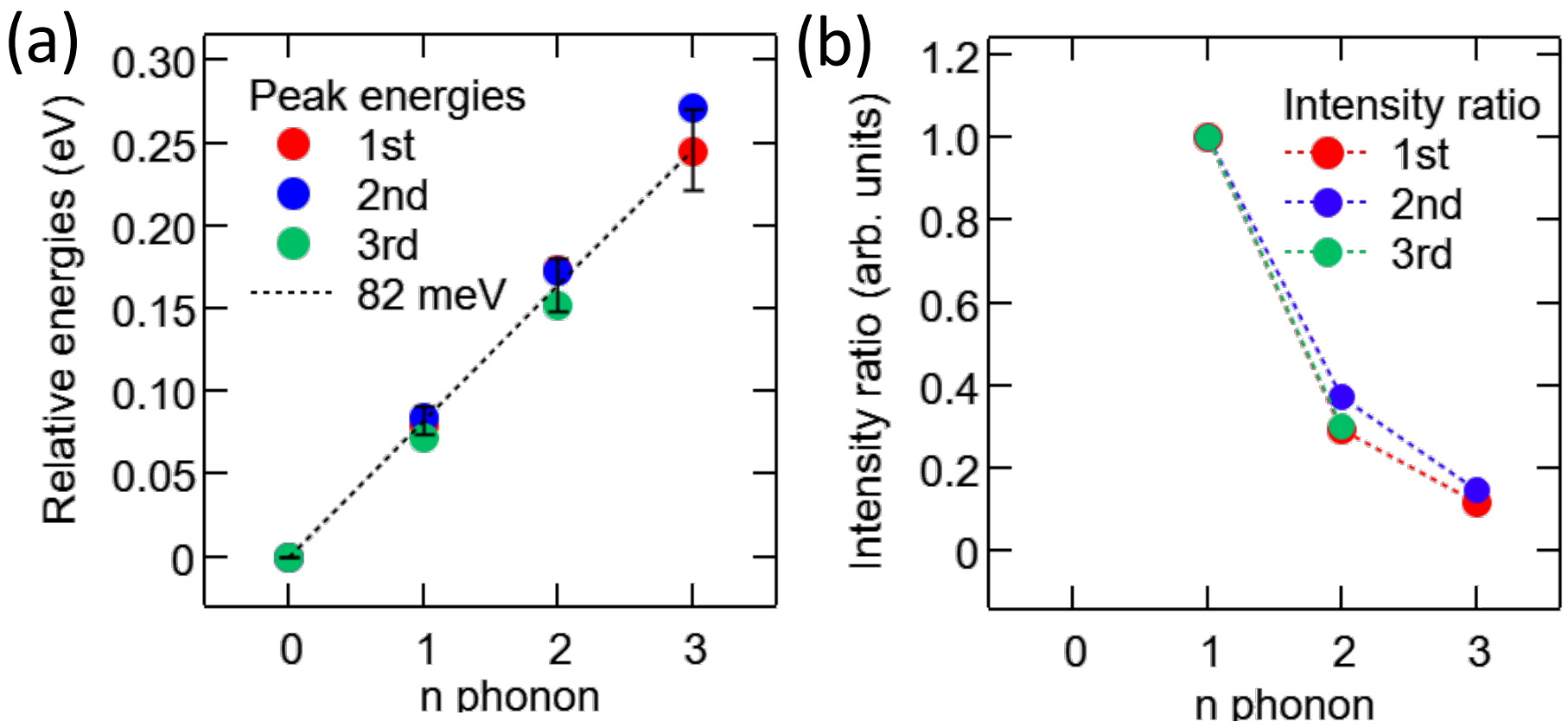

Supplementary Figure 1: Line-shape analysis of the phonon side bands in the RIXS spectrum in Fig. 4(a), (a) Relative energies of the phonon satellites. (b) Relative intensities of the phonon satellites. 1st: $J_{\mathrm{eff}} = 2 \rightarrow 2$ multiplet excitation, 2nd: $J_{\mathrm{eff}} = 2 \rightarrow 0, 1$ excitation, 3rd: $J_{\mathrm{eff}} = 2 \rightarrow 2$ excitation.

Supplementary Table 1: Parameter values of the Voigt functions for the three multiplet excitation peaks.

| Parameter | 1st | 2nd | 3rd |
|---|---|---|---|
| Peak energy $\omega_0$ (eV) | 0.0 | 0.397 | 0.788 |
| FWHM of Gaussian $\Gamma_G$ (meV) | 32.4 | 45.2 | 45.0 |
| Integrated intensity (arb. units) | 216.6 | 145.1 | 36.0 |
| FWHM ratio $\Gamma_L/\Gamma_G$ | 0.85 | 0.54 | 1.28 |

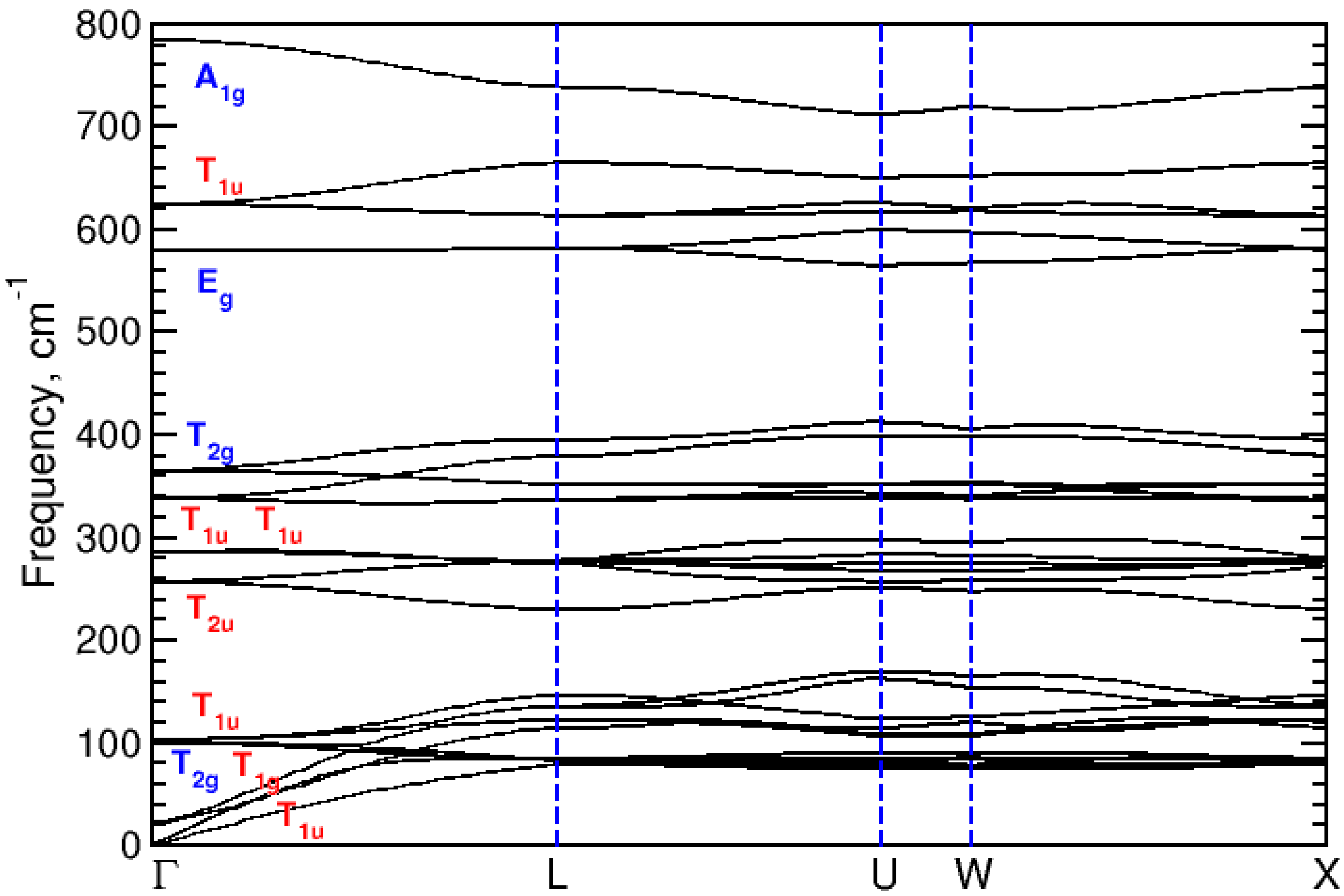

Supplementary Figure 2: Phonon dispersions (together with characters) as calculated by non-magnetic DFT.

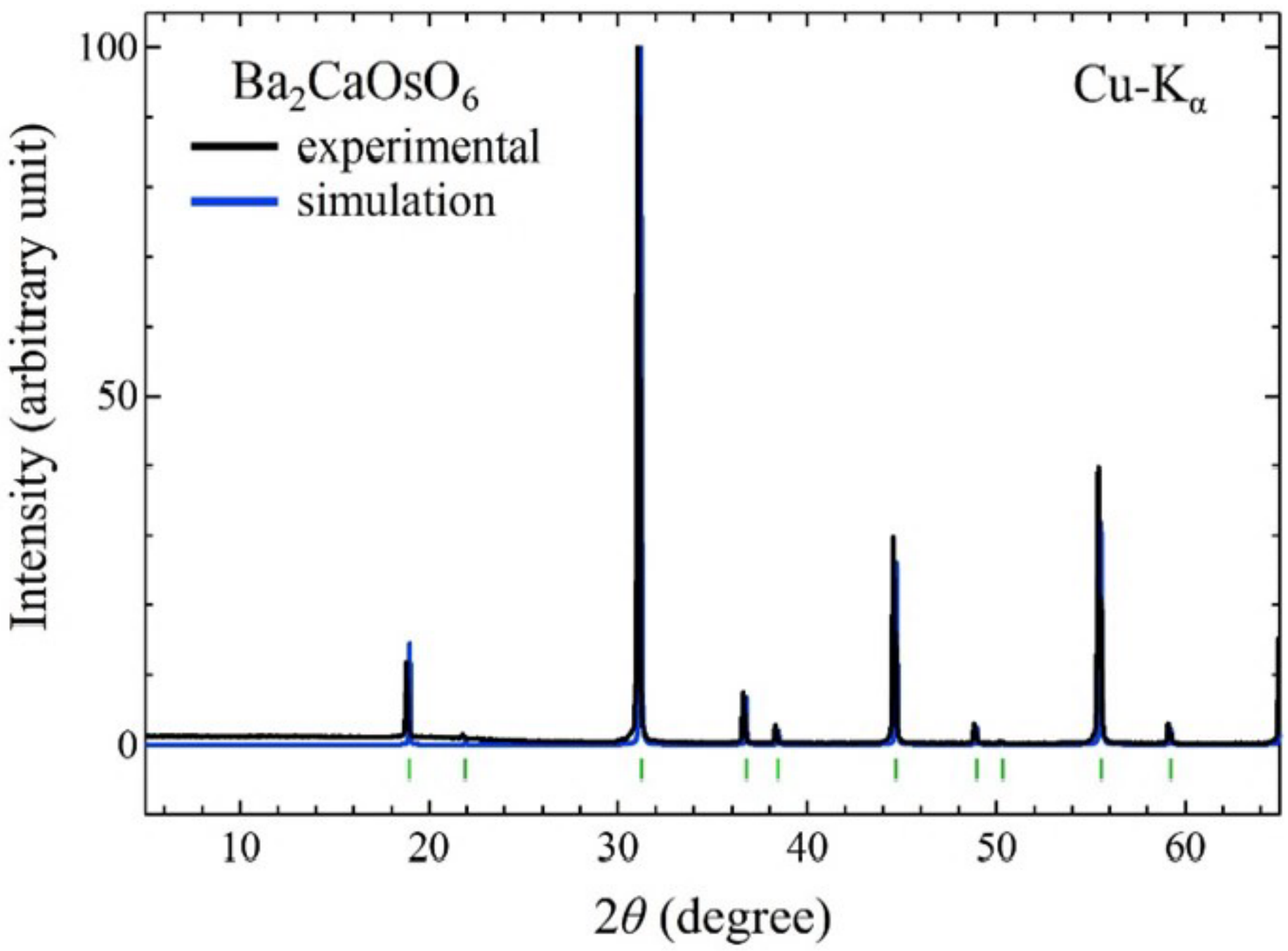

Supplementary Figure 3: Powder X-ray diffraction of the polycrystalline $Ba_2CaOsO_6$ sample.